\documentclass[sigconf,natbib=false]{acmart}
\AtBeginDocument{%
  }

\makeatletter
\renewcommand{\@copyrightpermission}{}
\makeatother
\setcopyright{acmlicensed}
\copyrightyear{2026}
\acmYear{2026}
\acmDOI{XXXXXXX.XXXXXXX}
\acmConference[SIGIR '26]{SIGIR}{July 20--24,
  2026}{Melbourne | Naarm, Australia}
\RequirePackage[
  datamodel=acmdatamodel,
  style=acmnumeric,
  ]{biblatex}

\newcommand{\ignore}[1]{}
\usepackage{subcaption}
\begin{document}

\title{ADaFuSE: Adaptive Diffusion-generated Image and Text Fusion for Interactive Text-to-Image Retrieval}

\author{Zhuocheng Zhang}
\affiliation{
  \institution{Hunan University}
  \city{Changsha}
  \state{Hunan}
  \country{China}
}
\email{zhuocheng@hnu.edu.cn}

\author{Xingwu Zhang}
\affiliation{
  \institution{Hunan University}
  \city{Changsha}
  \state{Hunan}
  \country{China}
}
\email{xingwuzhang@hnu.edu.cn}

\author{Kangheng Liang}
\affiliation{
  \institution{University of Glasgow}
  \city{Glasgow}
  \country{United Kingdom}
}
\email{2743944l@student.gla.ac.uk}

\author{Guanxuan Li}
\affiliation{
  \institution{Hunan University}
  \city{Changsha}
  \state{Hunan}
  \country{China}
}
\email{1005890623lgx@gmail.com}

\author{Richard Mccreadie}
\affiliation{
  \institution{University of Glasgow}
  \city{Glasgow}
  \country{United Kingdom}
}
\email{Richard.Mccreadie@glasgow.ac.uk}

\author{Zijun Long}
\authornote{Corresponding author.}
\affiliation{
  \institution{Hunan University}
  \city{Changsha}
  \state{Hunan}
  \country{China}
}
\email{longzijun@hnu.edu.cn}

\renewcommand{\shortauthors}{Trovato et al.}
\newcommand{\todo}[1]{\textcolor{red}{#1}}

\definecolor{lightgreen}{rgb}{0.0,0.8,0.3}

\newcommand{\zc}[1]{\textcolor{black}{#1}}
\newcommand{\ric}[1]{\textcolor{black}{#1}}

\newcommand{\myheader}[1]{\vspace{0.4em}\noindent\textbf{#1:}}
\begin{abstract}
Recent advances in interactive text-to-image retrieval (I-TIR) use diffusion models to bridge the modality gap between the textual information need and the images to be searched, resulting in increased effectiveness. However, existing frameworks fuse multi-modal views of user feedback by simple embedding addition. In this work, we show that this static and undifferentiated fusion indiscriminately incorporates generative noise produced by the diffusion model, leading to performance degradation for up to 55.62\% samples. We further propose ADaFuSE (Adaptive Diffusion-Text Fusion with Semantic-aware Experts), a lightweight fusion model designed to align and calibrate multi-modal views for diffusion-augmented I-TIR, which can be plugged into existing frameworks without modifying the backbone encoder. Specifically, we introduce a dual-branch fusion mechanism that employs an adaptive gating branch to dynamically balance modality reliability, alongside a semantic-aware mixture-of-experts branch to capture fine-grained cross-modal nuances.
Via thorough evaluation over four standard I‑TIR benchmarks, ADaFuSE achieves state-of-the-art performance, surpassing DAR by up to 3.49\% in Hits@10 with only a 5.29\% parameter increase, while exhibiting stronger robustness to noisy and longer interactive queries. These results show that generative augmentation coupled with principled fusion provides a simple, generalizable alternative to fine‑tuning for interactive retrieval.
\end{abstract}


%\ccsdesc[500]{Do Not Use This Code~Generate the Correct Terms for Your Paper}
%\ccsdesc[300]{Do Not Use This Code~Generate the Correct Terms for Your Paper}
%\ccsdesc{Do Not Use This Code~Generate the Correct Terms for Your Paper}
%\ccsdesc[100]{Do Not Use This Code~Generate the Correct Terms for Your Paper}


\maketitle

\section{Introduction}

Interactive text-to-image retrieval (I-TIR) allows users to find target images in a corpus by iteratively incorporating user feedback expressed as natural language dialogue~\cite{guo2018dialog, Lee2024_PlugIR,Levy2023_ChatIR}. 
Recently, diffusion-augmented I-TIR has emerged as a compelling paradigm by employing diffusion models~\cite{ho2020denoising, saharia2022photorealistic} to generate synthetic images conditioned on the dialogue context as visual proxies, serving to enrich text queries~\cite{long2025diffusion} or act as standalone image queries~\cite{Yang2025_GenIR}. Existing frameworks, such as DAR~\cite{long2025diffusion}, integrate the dialogue text and corresponding generated images via a static additive fusion strategy, combining their embeddings with a fixed weight.

However, we argue that this static additive fusion suffers from two key limitations. First, methods like DAR that rely on fixed weighting ignore how the usefulness of each modality varies from instance-to-instance. In practice, the reliance on visual and textual information should be dynamic, depending on how well the generated image captures the current intent~\cite{arevalo2017gated, zhang2023provable, alfasly2022learnable}. Second, static additive fusion treats all generated images as equally valuable, ignoring their actual quality. Since diffusion models are inherently stochastic, the consistency between generated images and user intent fluctuates across samples. Static addition indiscriminately fuses these generated images, inevitably introducing noise.

To address the above limitations, we propose ADaFuSE (Adaptive Diffusion-Text Fusion with Semantic-aware Experts), a lightweight fusion model that is designed to dynamically calibrate multi-modal representations for diffusion-augmented I-TIR, which can be plugged into existing I-TIR pipelines without modifying the backbone encoder. Structurally, the model comprises two coordinated branches: an adaptive gating~\cite{perez2018film} branch that leverages cross-modal interaction to dynamically modulate the fusion weights of text features and corresponding generated image features; and a semantic-aware mixture-of-experts branch that utilizes diverse semantic-aware experts to construct compensatory features, capturing fine-grained cross-modal nuances. By integrating the modulated features from the gating branch with the compensatory features from the semantic-aware branch, ADaFuSE forms a more robust and intent-aligned query representation than static additive fusion.

Our main contributions are: (i) We critically analyze existing diffusion-augmented I-TIR frameworks, exposing the limitations of undifferentiated static additive fusion.
(ii) We propose ADaFuSE, a lightweight fusion model designed to achieve a robust fused representation between input text and diffusion-generated images. (iii) We demonstrate state‑of‑the‑art performance on four standard I‑TIR benchmarks, improving over strongest diffusion‑augmented baseline (DAR)~\cite{long2025diffusion} by up to 3.49\% in Hits@10 and showing robustness to increasing query complexity and interaction length.

\section{Related Work}
\myheader{Interactive Text-to-Image Retrieval}
Unlike traditional retrieval systems that rely solely on a single (short) query~\cite{lee2018stacked, radford2021learning}, interactive text-to-image retrieval (I-TIR) iteratively refines a search query based on multi-round user feedback~\cite{Levy2023_ChatIR,zhu2024enhancing}. Driven by advancements in large language models~\cite{brown2020language, grattafiori2024llama, liu2024deepseek} and vision language models~\cite{lu2019vilbert, radford2021learning, li2022blip}, this interactive approach has been used to increase search effectiveness for a wide range of use-cases, ranging from general image search~\cite{Lee2024_PlugIR,Yang2025_GenIR,zhao2025chatsearch} to specialized tasks like person retrieval~\cite{bai2025chat,lu2025llava}. Diffusion-augmented I-TIR approaches are a recent innovation that incorporates diffusion-generated images to bridge the semantic gap between the textual query/dialogue and the images being searched. Specifically, DAR~\cite{long2025diffusion} synthesizes these images conditioned on the dialogue context and integrates them with textual queries via a static additive fusion strategy, enabling state-of-the-art performance, even in zero-shot settings.

\looseness -1 \myheader{Fusion of Multi-modal Query Views} To our knowledge, no prior work examines how to better fuse multi-modal query views for diffusion-augmented I-TIR. The closest related field is composed image retrieval (CIR)~\cite{vo2019composing, wu2021fashion, liu2021image, anwaar2021compositional}, where a multi-modal query, typically a reference image paired with a modification text, is used to retrieve the target image~\cite{song2025comprehensive,zhang2025composed}. This task has been extensively explored in domains ranging from fashion e-commerce to open-domain scenes, driven by large-scale datasets such as FashionIQ~\cite{wu2021fashion}, CIRR~\cite{liu2021image} and CIRCO~\cite{baldrati2023zero}. However, diffusion-augmented I-TIR differs from CIR in the nature of its inputs. In standard CIR, the text typically acts as a modification instruction applied to a reliable reference image. Conversely, diffusion-augmented retrieval employs a synthetic image generated from (potentially long) dialogue context~\cite{long2025diffusion,Yang2025_GenIR}. In this setting, theoretically both modalities should convey the same semantic intent, but in practice the generated image will introduce instance-dependent noise~\cite{zhang2026eliminating}. This motivates the development of dedicated fusion mechanisms tailored to diffusion-augmented I-TIR.

\begin{figure}[h]
    \centering
    \begin{subfigure}{1.0\linewidth}
        \centering \includegraphics[width=1.0\linewidth]{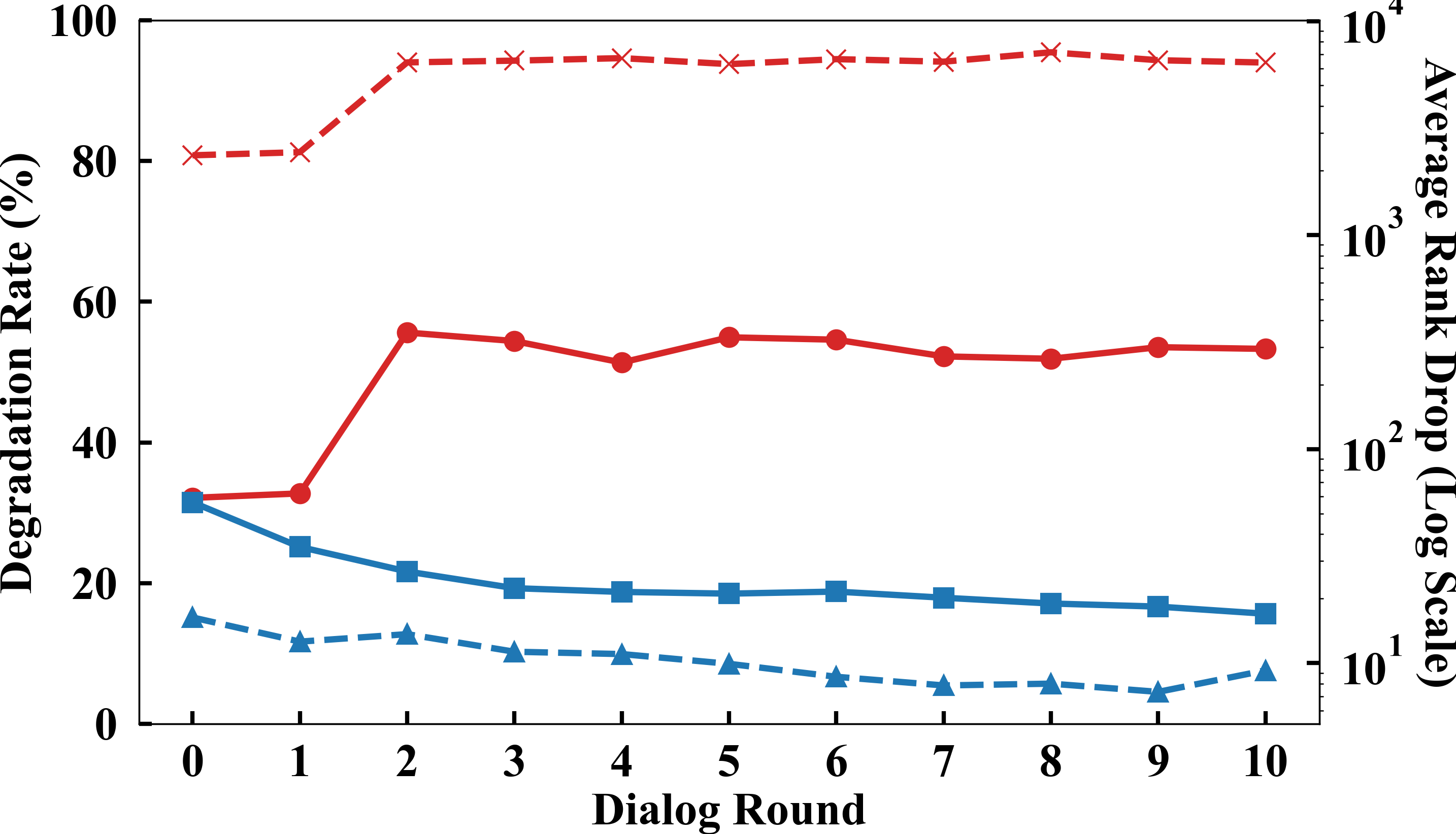}
    \end{subfigure}

    \begin{subfigure}{1.0\linewidth}
        \centering  \includegraphics[width=1.0\linewidth]{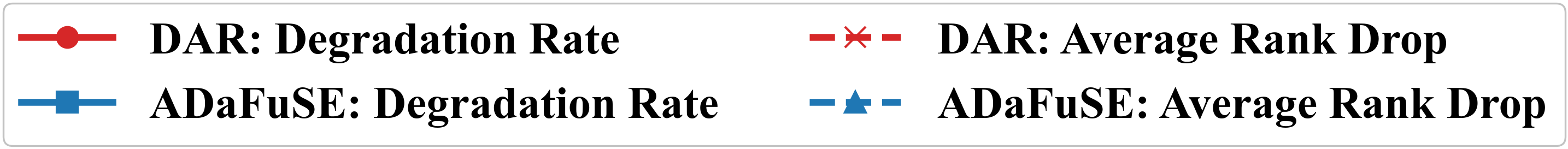}
    \end{subfigure}
    \vspace{-0.3cm} 
    \caption{Analysis of the diffusion-augmented degradation rate (left) and average rank drop (right) on VisDial\cite{das2017visual} validation set. Lower is better.}
    \label{fig:adr_metric}
    \vspace{-5mm}
\end{figure}

\section{Proposed Method: ADaFuSE}
\subsection{Limitations of Additive Fusion}
\label{sec:limits}
\looseness-2
As discussed in \cite{zhang2026eliminating}, generated images produced by diffusion-augmented interactive text-to-image retrieval (I-TIR) approaches often semantically deviate from the user's retrieval intent, introducing noise that can conflict with the original dialogue context. The state-of-the-art method, DAR~\cite{long2025diffusion}, uses a static weighted additive fusion strategy, combining these diffusion-generated images with the query representation directly. While DAR has been demonstrated to be effective regardless of this limitation, we argue that the introduced noise is degrading performance for some queries.

To quantify this risk, we analyze the proportion of queries where incorporating a diffusion-generated image hurts rather than improves the retrieval performance (referred to as the degradation rate), as well as average rank drop of the relevant image. As illustrated in Figure~\ref{fig:adr_metric}, the red solid line reveals that DAR results in a degradation rate exceeding 50\% starting from round 2, accompanied by a large average rank drop for degraded queries of approximately 7500, as indicated by the red dashed line. This demonstrates that diffusion noise is a major issue, and that a better way to fuse the text and diffusion-generated image evidence is needed.

\begin{figure*}[htp]
    \centering
    \setlength{\abovecaptionskip}{3pt}
    \setlength{\belowcaptionskip}{0pt}
    \includegraphics[width=0.9\linewidth]{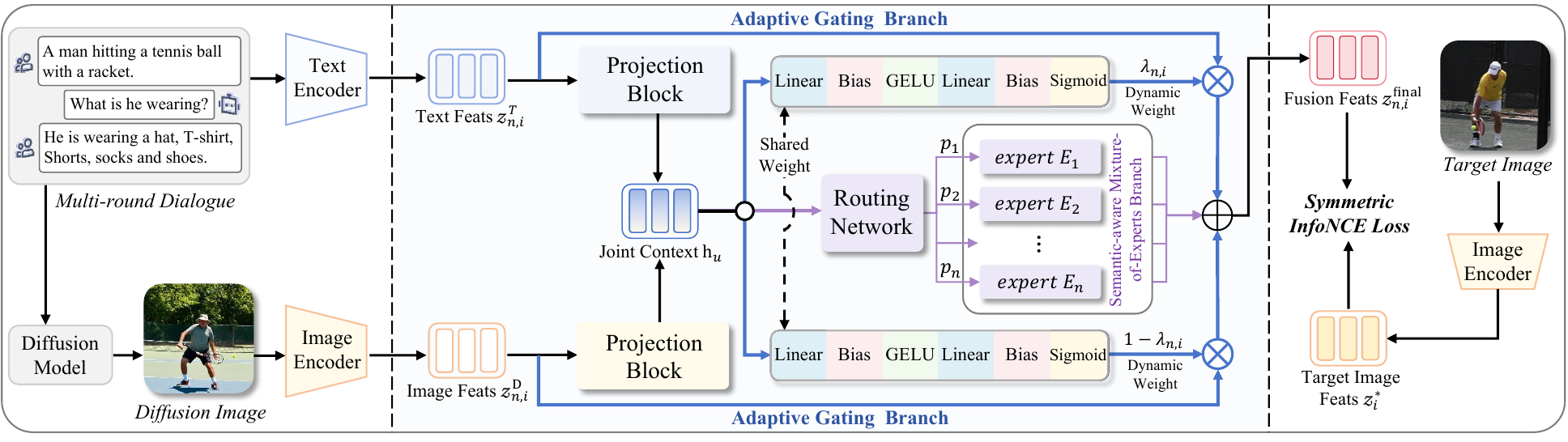}
    %\vspace{-6mm}
    \caption{The Architecture overview of ADaFuSE.}
    \label{fig:Figure_fw}
    \vspace{-1mm}
\end{figure*}
\subsection{Adaptive Fusion of Diffusion-generated} Images and Text
We hypothesize that an effective mean to limit the leaking of diffusion-generated noise into the ranking process is to be more selective when fusing the text and diffusion-produced evidence. To achieve this, we propose ADaFuSE (Adaptive Diffusion-Text Fusion with Semantic-aware Experts), \zc{as illustrated in Figure~\ref{fig:Figure_fw},} a lightweight model that acts as a smart bridge between the text and image modalities, dynamically calibrating how much signal from each modality to use per-query.

\myheader{Query Encoding and Projection} 
Consider the $i$-th sample in the dataset at dialogue round $n$. We denote the textual query as $T_{n,i}$, the corresponding diffusion-generated image as $I_{n,i}$, and the ground-truth target image as $I^*_i$. Let $\Phi_T(\cdot)$ and $\Phi_I(\cdot)$ denote the text and image encoder, respectively. The initial embeddings are obtained by mapping inputs into a shared $d$-dimensional embedding space:
\begin{equation}
z_{n,i}^T = \Phi_T(T_{n,i}), \quad z_{n,i}^D = \Phi_I(I_{n,i}), \quad z^*_i = \Phi_I(I^*_i)
\end{equation}
where $z_{n,i}^T, z_{n,i}^D, z^*_i \in \mathbb{R}^d$. 
Although pre-trained encoders effectively align modalities globally, their pre-training objectives prioritize invariance, which can suppress fine-grained visual details containing valuable semantic information for supplementing the text~\cite{chen2020simple,chen2021exploring}. Direct fusion within this compressed space restricts the effective utilization of these visual cues. To recover this lost capacity, ADaFuSE first employs two projection blocks (see Figure~\ref{fig:Figure_fw}) to non-linearly project the raw text and image embeddings into a higher-dimensional, task-specific latent space:
\begin{equation}
  % \mathbf{h}_{n,i}^T = \mathrm{GELU}\!\big(\mathcal{P}_T(z_{n,i}^T)\big), \quad
  % \mathbf{h}_{n,i}^D = \mathrm{GELU}\!\big(\mathcal{P}_D(z_{n,i}^D)\big)
    \mathbf{h}_{n,i}^T = \delta\big(\mathcal{P}_T(z_{n,i}^T)\big), \quad
    \mathbf{h}_{n,i}^D = \delta\big(\mathcal{P}_D(z_{n,i}^D)\big)
\end{equation}
where $\mathcal{P}_{\{T,D\}}(\cdot): \mathbb{R}^d \rightarrow \mathbb{R}^{d'}$ are instantiated as independent projection heads to capture modality-specific characteristics and $\delta(\cdot)$ is the GELU activation function. This non-linear dimensional expansion serves to restore the discriminative capacity of the features, providing a more expressive embedding space for the subsequent \ignore{fine-grained conflict resolution}\zc{adaptive gating} and expert routing.

\myheader{Joint Context Initialization} To provide an initial joint representation of the textual query and the generated visual cues, we concatenate the projected features, denoted as $\mathbf{h}_{u}$:
\begin{equation}
    \mathbf{h}_{u} = [\mathbf{h}_{n,i}^T; \mathbf{h}_{n,i}^D] \in \mathbb{R}^{2d'}
\end{equation}
This unified representation $\mathbf{h}_{u}$ preserves the distinct characteristics of both modalities and serves as the fundamental input for the subsequent two branches in ADaFuSE.

\myheader{Adaptive Gating Branch}
In diffusion-augmented I-TIR, the generated image $I_{n,i}$ may contain details that are not fully consistent with the text~\cite{lim2025evaluating,zhang2026eliminating}, and we demonstrated in Section~\ref{sec:limits} that this can strongly degrade performance for some instances. On the other hand, these images do provide discriminative cues for many instances (e.g., fine-grained attributes or structural cues)~\cite{long2025diffusion}. Hence, we need a mechanism to distinguish to what extent each generated image is providing signal rather than noise. To address this, ADaFuSE employs an adaptive gating branch to estimate the reliability of the diffusion-generated image from the joint context $\mathbf{h}_{u}$ and outputs a dynamic scalar weight $\lambda_{n,i} \in (0, 1)$:  

%Conversely, $T_{n,i}$ is semantically stable but lacks visual specificity. 
%Static additive fusion with fixed weights ignores this instance-level reliability, potentially diluting critical retrieval features.

% \begin{equation}
%     \lambda_{n,i} = \sigma(\mathbf{W}_2 \cdot \delta(\mathbf{W}_1 \mathbf{h}_{u} + \mathbf{b}_1) + \mathbf{b}_2)
% \end{equation}
\begin{equation}
    \lambda_{n,i} = \sigma\big(\mathbf{W}_2 \cdot \delta(\mathbf{W}_1 \mathbf{h}_{u} + \mathbf{b}_1) + \mathbf{b}_2\big)
\end{equation}
where $\mathbf{W}_1 \in \mathbb{R}^{d_{mid} \times 2d'}$ and $\mathbf{W}_2 \in \mathbb{R}^{1 \times d_{mid}}$ are learnable weights, $\mathbf{b}_1, \mathbf{b}_2$ are biases, and $\Large\sigma(\cdot)$ is the Sigmoid function. This gate $\lambda_{n,i}$ modulates the fusion of raw semantic embeddings:
\begin{equation}
    \label{lambda}
    \mathbf{z}_{n,i}^{base} = \lambda_{n,i} \cdot z_{n,i}^T + (1 - \lambda_{n,i}) \cdot z_{n,i}^D
\end{equation}

Intuitively, as the divergence between a textual query and associated diffusion generated image increases, retrieval performance will degrade, since the image will pull the final joint embedding toward visually salient yet semantically incorrect cues, overwhelming the anchoring effect of the textual signal. ADaFuSE counteracts this by predicting, from the joint context $\mathbf{h}_u$, how much the system should trust the diffusion-derived cues for each dialogue round and query. As a result, ADaFuSE can selectively exploit the generated visual signal when $I_{n,i}$ is predicted to be reliable, while falling back to the text representation when the generated image is noisy or contradictory; thereby avoiding the brittleness of fixed-weight fusion and yielding more robust alignment with user intent.

\myheader{Semantic-aware Mixture-of-Experts Branch} While the adaptive gating provides a low-cost means \zc{to} limit the impact \zc{of} semantic mismatches between the \zc{text} \ignore{query}and generated image, it is a \ignore{course}\zc{coarse}-grained re-weighting of the two modalities. Not all semantic divergencies will be problematic; the diffusion model can probabilistically produce image elements that initially seem irrelevant, but that are still useful during retrieval. ADaFuSE learns to correct for this using a semantic-aware mixture-of-experts (MoE)~\cite{fedus2022switch, han2024fusemoe} branch. We instantiate a set of $K$ semantic-aware experts $\{E_k\}_{k=1}^K$. Each expert $E_k(\cdot): \mathbb{R}^{2d'} \to \mathbb{R}^{d_{hidden}}$ is implemented as a lightweight feed-forward network with independent learnable parameters. A routing network is applied to estimate the relevance of each expert conditioned on the current joint context $\mathbf{h}_{u}$, producing a routing probability\ignore{distribution} $p_k$ \zc{for the $k$-th expert :
\begin{equation}
p_k = \frac{\exp(\mathcal{R}_k(\mathbf{h}_u))}{\sum_{j=1}^{K} \exp(\mathcal{R}_j(\mathbf{h}_u))}
\end{equation}
where $\mathcal{R}_k(\cdot)$ denotes the $k$-th logit output of the routing network, designed as a bottleneck two-layer MLP.} Subsequently, the outputs of the experts are aggregated via weighted addition to form a compensatory feature vector $\mathbf{h}_{res}$:
\begin{equation}
    \mathbf{h}_{res} = \sum_{k=1}^K p_k E_k(\mathbf{h}_{u})
\end{equation}
Finally, we project the MoE output back to the original embedding space dimension $d$ using a projection matrix
$\mathbf{W}_{out} \in \mathbb{R}^{d \times d_{hidden}}$. 
This projection is added to the gating-based fusion output as a residual term to obtain the final ADaFuSE query representation, followed by $\ell_2$ normalization:
\begin{equation}
    \mathbf{z}_{n,i}^{final} = \text{Normalize}(\mathbf{z}_{n,i}^{base} + \mathbf{W}_{out}\mathbf{h}_{res})
\end{equation}
% This residual design enables the model to enrich the linear base representation with mined non-linear semantic details, yielding a robust and precise query representation.

Unlike the adaptive gating branch, which can only interpolate between two existing modality embeddings, the semantic-aware MoE branch provides additional capacity to model higher-order cross-modal interactions that are not expressible through linear re-weighting. In diffusion-augmented I-TIR, the useful signal often lies in subtle compositional cues (e.g., “shorter sleeves but same silhouette”), which require non-linear feature synthesis rather than selecting one modality over the other. By routing each instance to a small set of specialized experts, the MoE learns a conditional set of refinements tailored to the current dialogue context, while the residual connection preserves the base query to prevent overcorrection. Consequently, the final representation is selectively enriched with context-dependent, fine-grained semantics from $\mathbf{h}_{res}$, improving alignment with user intent.

\begin{figure*}[t]
    \centering
    % --- 第一行：图片部分 ---
    \begin{minipage}[b]{0.22\textwidth}
        \centering
        \includegraphics[width=\linewidth]{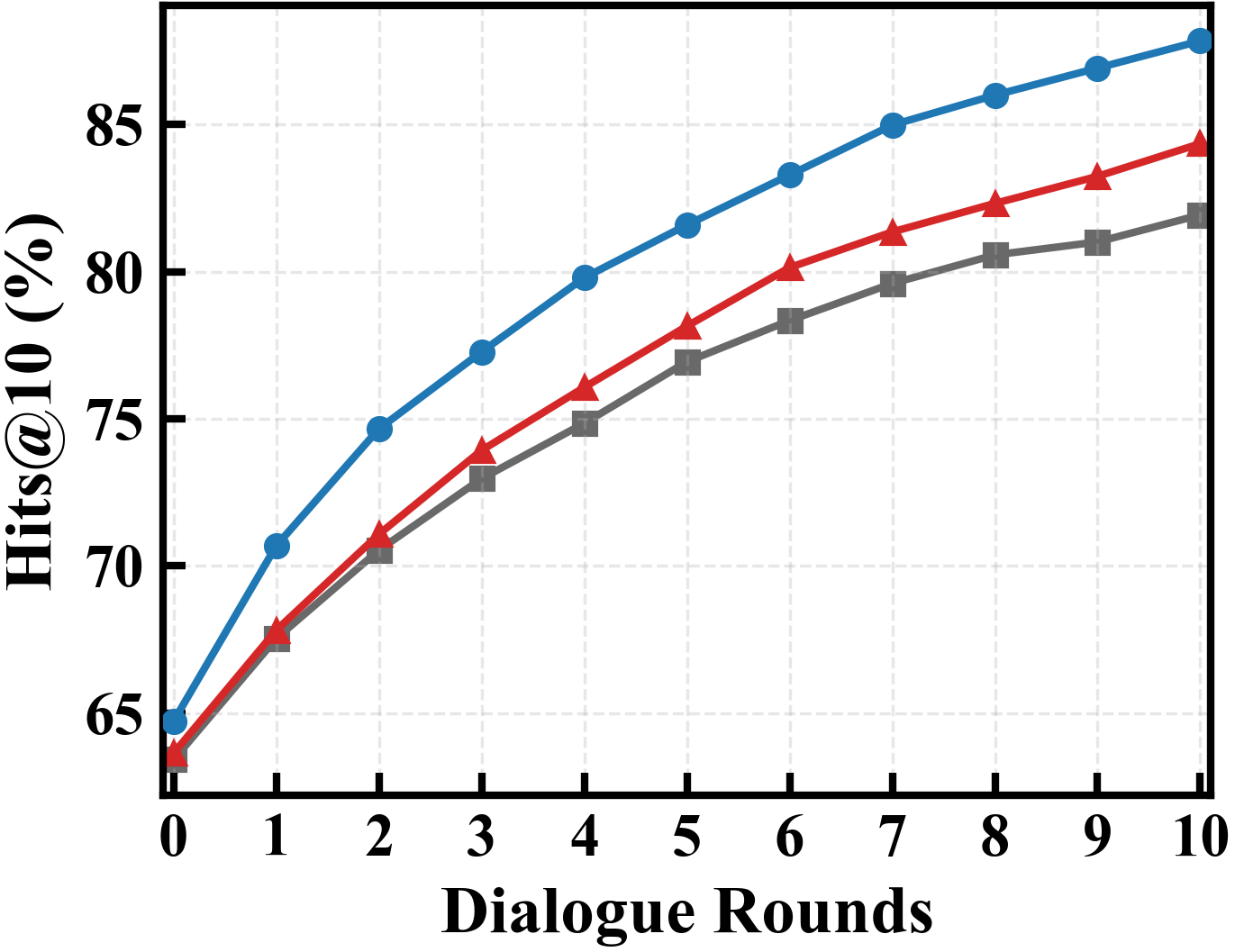}
        \vspace{-2mm} % 微调图片底部空白
    \end{minipage}
    \begin{minipage}[b]{0.22\textwidth}
        \centering
        \includegraphics[width=\linewidth]{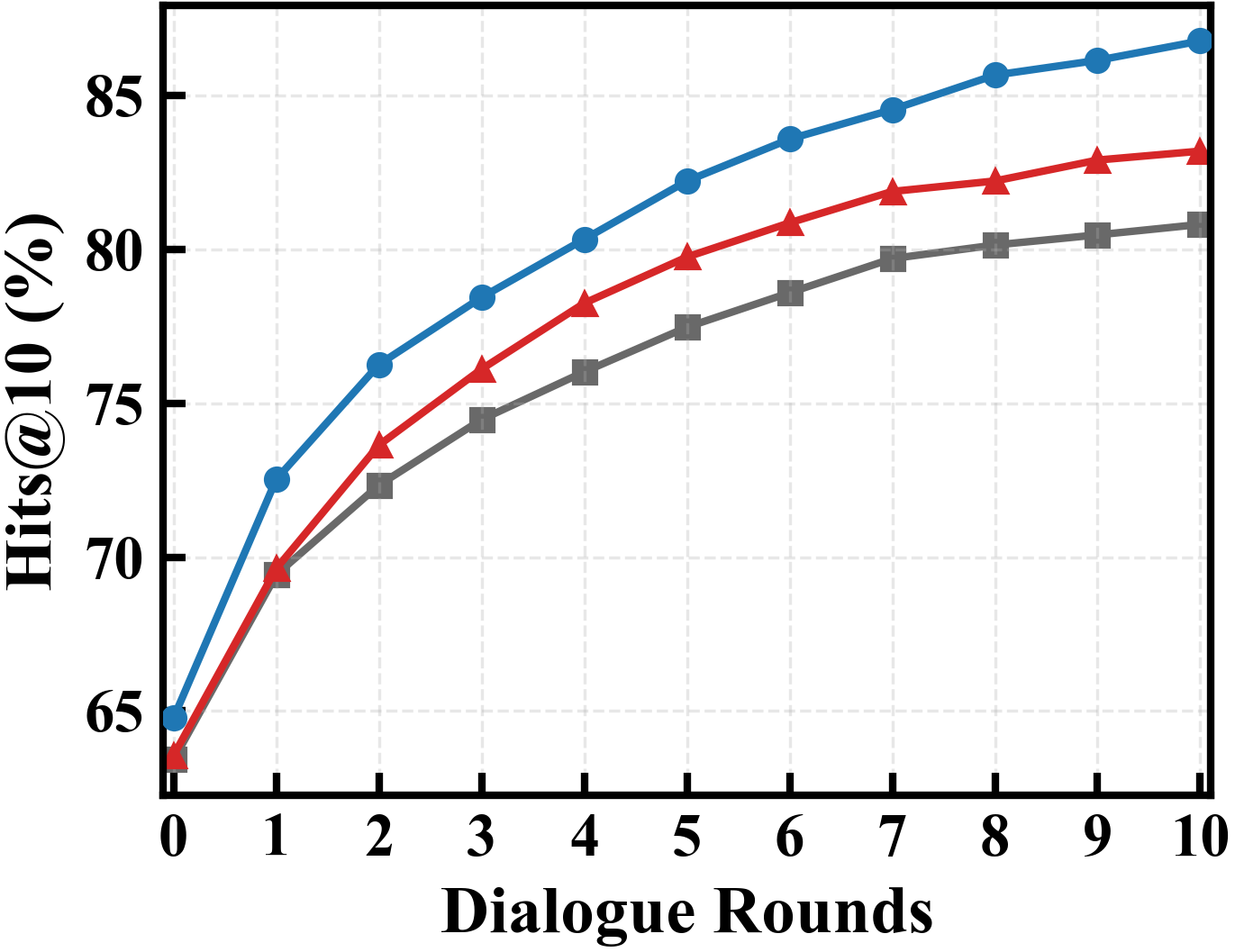}
        \vspace{-2mm}
    \end{minipage}
    \begin{minipage}[b]{0.22\textwidth}
        \centering
        \includegraphics[width=\linewidth]{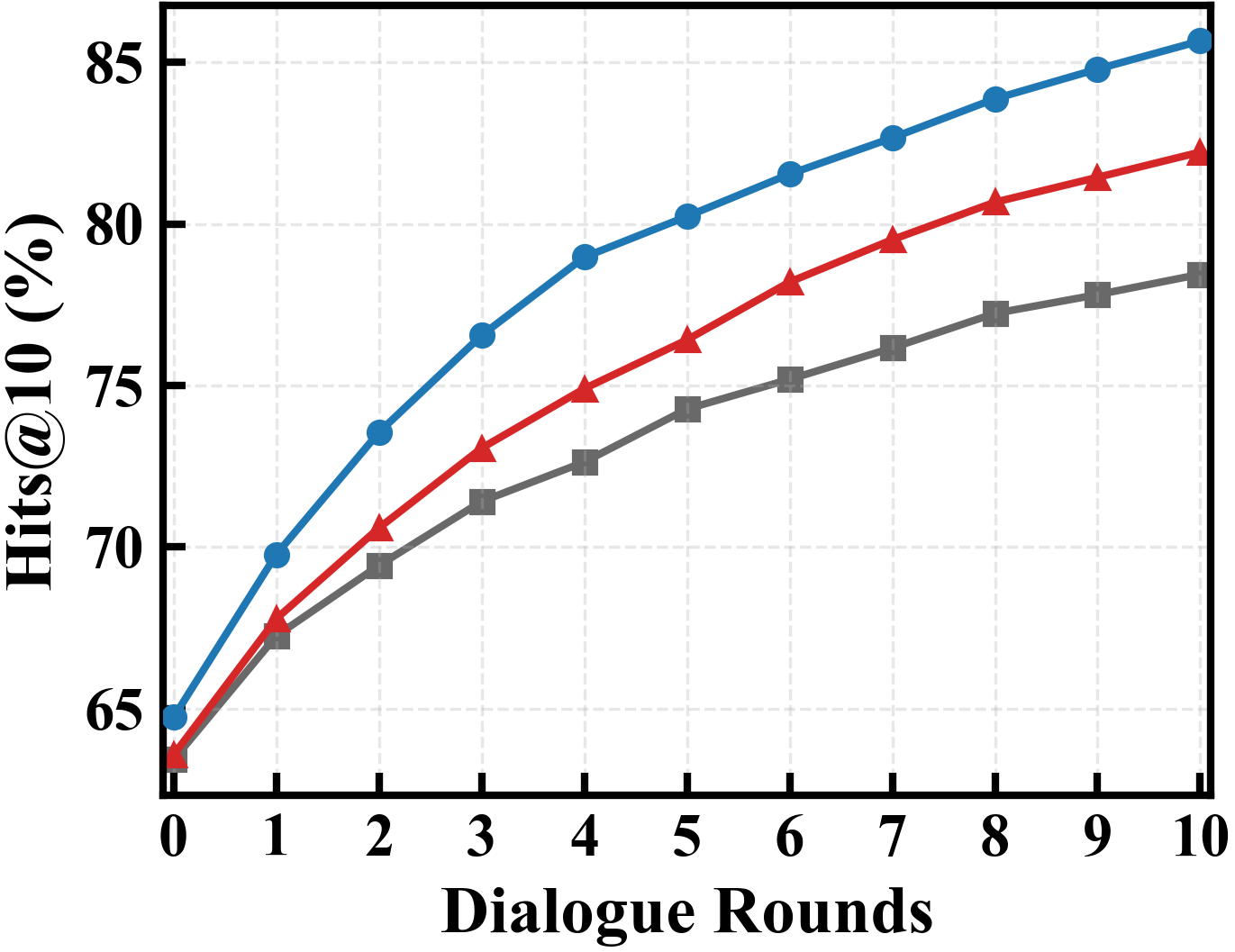}
        \vspace{-2mm}
    \end{minipage}
    \begin{minipage}[b]{0.22\textwidth}
        \centering
        \includegraphics[width=\linewidth]{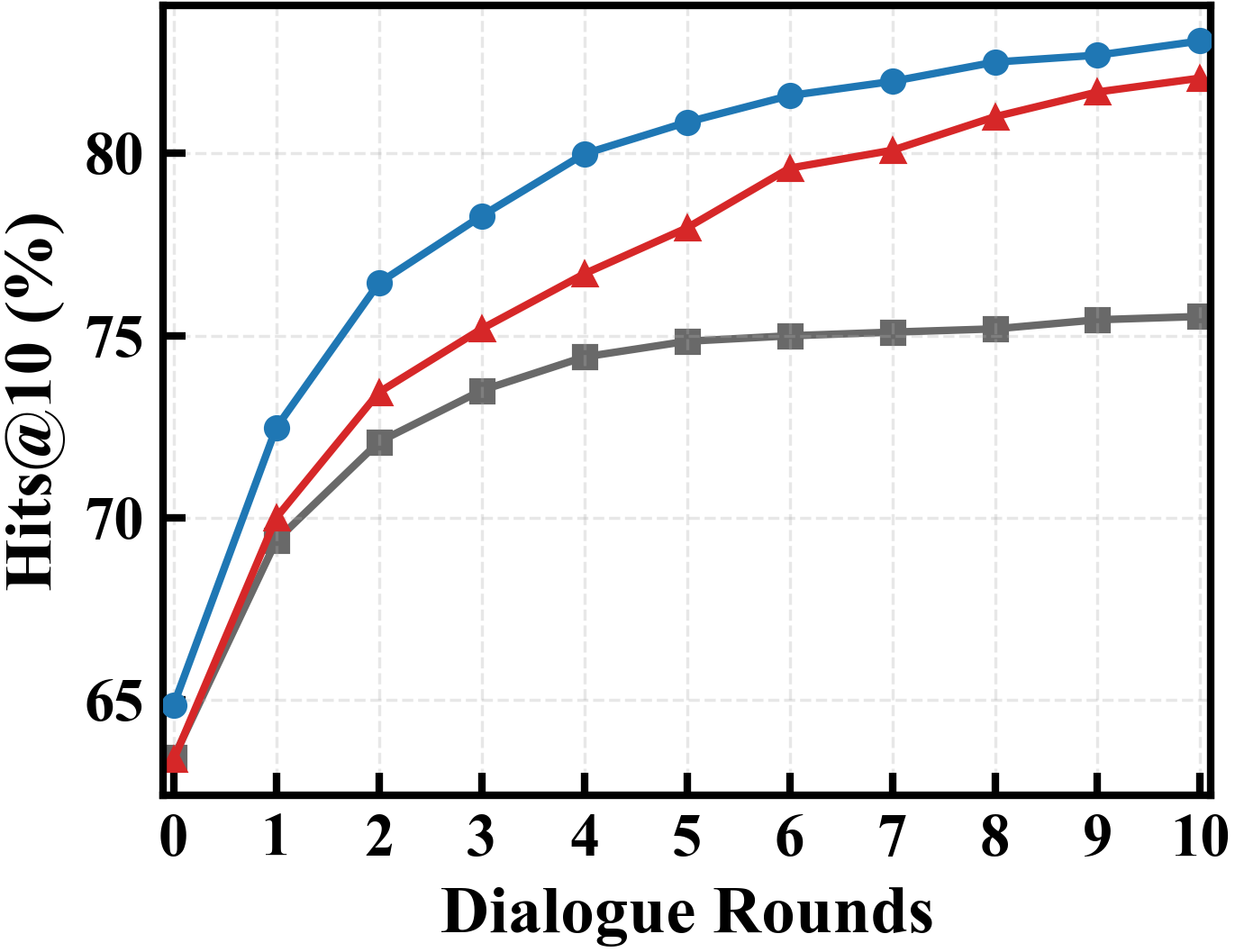}
        \vspace{-2mm}
    \end{minipage}

    % --- 第二行：图例部分 (放在图片和标题中间) ---
    \vspace{-2mm} % 图片与图例的间距
    \includegraphics[width=0.44\linewidth]{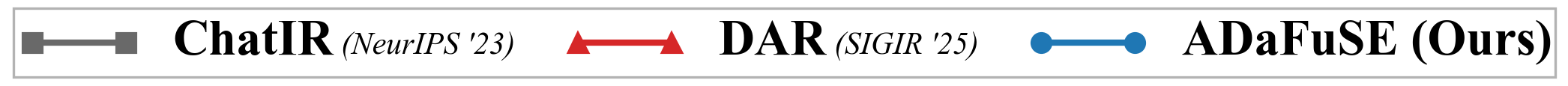}
    \vspace{-6mm} % 图例与下方标题的间距

    % --- 第三行：副标题部分 ---
    \par % 强制换行
    \hspace{6.5mm}
    \begin{minipage}[t]{0.22\textwidth}
        \centering
        \subcaption{VisDial} 
        \label{fig:res_visdial}
    \end{minipage}
    \hspace{0mm}
    \begin{minipage}[t]{0.22\textwidth}
        \centering
        \subcaption{ChatGPT\_BLIP2}
        \label{fig:res_chatgpt}
    \end{minipage}
    \hspace{0mm}
    \begin{minipage}[t]{0.22\textwidth}
        \centering
        \subcaption{Human\_BLIP2}
        \label{fig:res_human}
    \end{minipage}
    \hspace{0mm}
    \begin{minipage}[t]{0.22\textwidth}
        \centering
        \subcaption{Flan-Alpaca-XXL\_BLIP2}
        \label{fig:res_flan}
    \end{minipage}
    \vspace{-4mm} 
    \caption{The overall performance (measured by Hits@10, the higher is better) across four benchmarks.}
    \label{fig:performance}
\end{figure*}

%Our proposed ADaFuSE consistently outperforms baselines across all dialogue rounds.

% \subsection{Training Objective} 

\section{Experiments and Analysis}

\subsection{Experimental Setup} We conduct our training on the DA-VisDial dataset~\cite{zhang2026eliminating}. Consistent with prior SOTA approaches~\cite{Levy2023_ChatIR, long2025diffusion}, we utilize the pre-trained BLIP weights~\cite{Levy2023_ChatIR} to initialize our encoder and optimize ADaFuSE in an end-to-end manner \zc{via the symmetric InfoNCE loss~\cite{radford2021learning, oord2018representation}}.
We evaluate our method on the VisDial~\cite{das2017visual} validation set and three additional I-TIR benchmarks introduced by ChatIR: \textit{ChatGPT\_BLIP2}, \textit{Human\_BLIP2}, \textit{Flan-Alpaca-XXL\_BLIP2}. We adopt the accumulated recall at rank 10 (Hits@10) as our primary evaluation metric.

We compare our proposed ADaFuSE model against two representative baselines built on the same ChatIR backbone: ChatIR and DAR.
(i) \textbf{ChatIR}~\cite{Levy2023_ChatIR}: a strong I-TIR baseline that relies solely on the dialogue text as the query.
(ii) \textbf{DAR}: a state-of-the-art diffusion-augmented variant that incorporates generated images, fusing the dialogue text and diffusion outputs via a static additive operation to refine the query representation.
For a fair comparison, all methods use the same pre-trained BLIP encoder as the base visual--textual backbone. The code used in this paper is publicly available at: 
\begin{center}
\underline{\emph{\url{https://anonymous.4open.science/r/ADaFuSE-E149/README.md}}}
\end{center}

% \subsection{Datasets and metrics}
% \todo{Simply illustrates the datasets we used for training and evaluation}

% We follow the standard experimental setting as in \todo{Cite chatir, plugir etc}. We employ the accumulated recall at rank K (Recall@K) as an evaluation metric, namely Hits@10 (H@10). Note that for each triplet, there is only a positive index image. Hence, each query has R@K zero or one. 

\subsection{Robustness across Rounds and Distributions}
\looseness -1 In this section, we investigate how ADaFuSE performs across dialogue rounds and data distributions, comparing it with state-of-the-art baselines on four benchmarks.
% \zc{Figure~\ref{fig:performance} compares the performance of ADaFuSE (blue line) against the state-of-the-art baselines across four benchmarks.}
As shown in Figure~\ref{fig:performance}a, on the in-distribution VisDial validation set, ADaFuSE consistently outperforms DAR, with Hits@10 gains growing from 1.09\% at round 0 to 3.49\% at round 10. This trend persists across all three out-of-distribution datasets, with ADaFuSE consistently outperforming the baselines at every interaction round (see Figure~\ref{fig:performance}b--d). We attribute the larger late-round gains to a pre-training mismatch; encoders trained on MSCOCO~\cite{lin2014microsoft}-style short captions handle early, brief dialogue well, but struggle as feedback becomes longer and may contain conflicting cues. ADaFuSE mitigates this by filtering out unreliable signals, leading to stronger improvements in later rounds.

Notably, Figure~\ref{fig:performance}d illustrates that the ChatIR baseline, which relies solely on textual queries, saturates early due to the dataset's prevalence of uninformative dialogue. While DAR mitigates this challenge by introducing diffusion-generated images, ADaFuSE achieves further improvements over DAR at every interaction round,
demonstrating that ADaFuSE is not only effective at fusion but is also capable of adaptively leveraging visual cues to compensate for low-quality textual feedback.
Moreover, since ADaFuSE and DAR share the same backbone encoder, DAR effectively serves as a non-adaptive fusion ablation, indicating that the observed performance gains stem from our adaptive fusion module rather than other components.

% Notably, on \textit{ChatGPT\_BLIP2} \zc{(see Figure~\ref{fig:performance}b)} and \textit{HUMAN\_BLIP2} \zc{ (see Figure~\ref{fig:performance}c) datasets}, our method expands its advantage, achieving a performance boost of 3.58\% and 3.44\% \ignore{at round 10}\zc{in Hits@10 after 10 dialog rounds} compared to the DAR baseline. 

\begin{figure}
    \centering
    \begin{subfigure}{0.85\linewidth}
        \centering \includegraphics[width=1.0\linewidth]{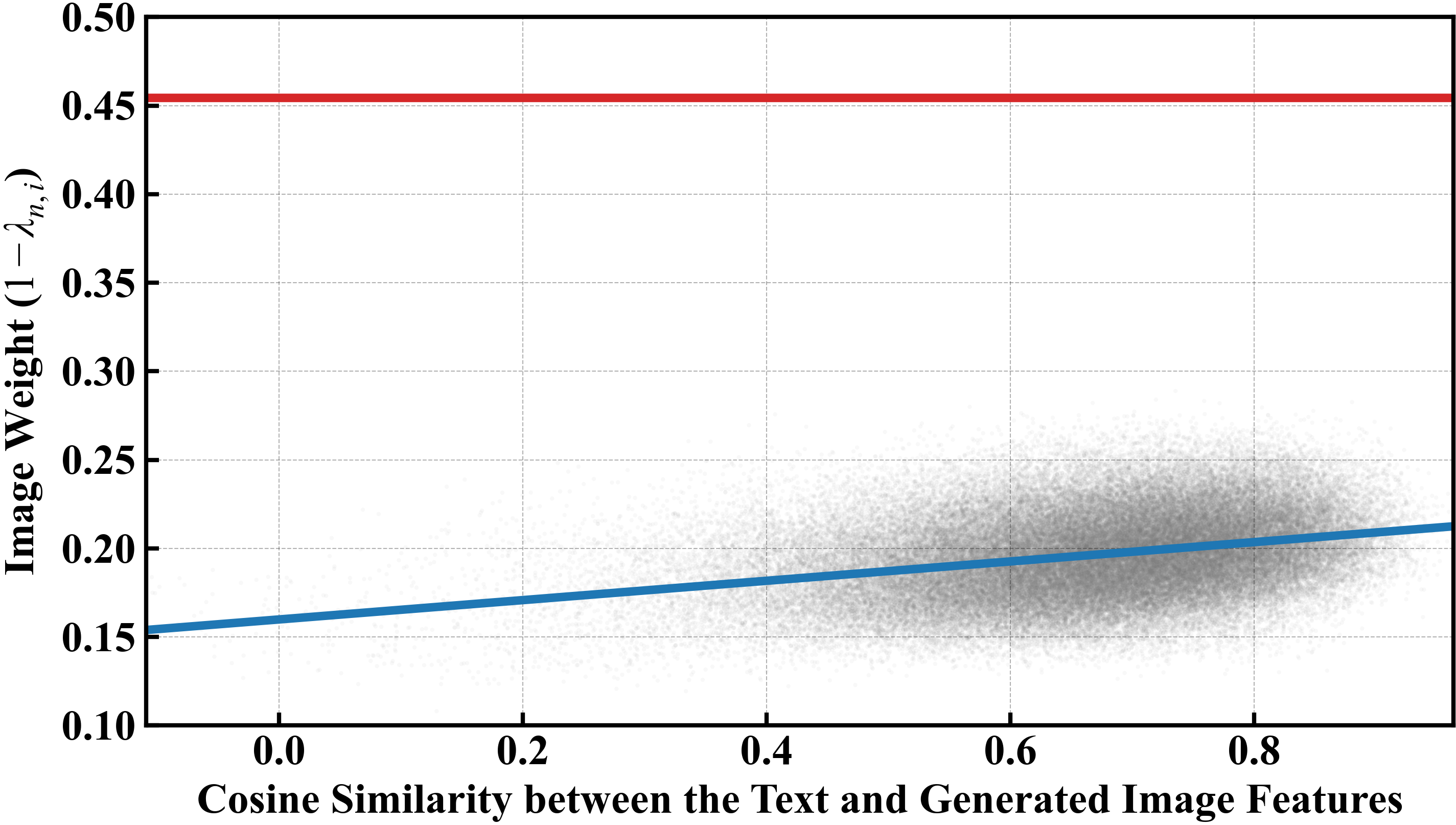}
    \end{subfigure}
    \vspace{-3.5mm} 
    \begin{subfigure}{1.0\linewidth}
        \centering  \includegraphics[width=1.0\linewidth]{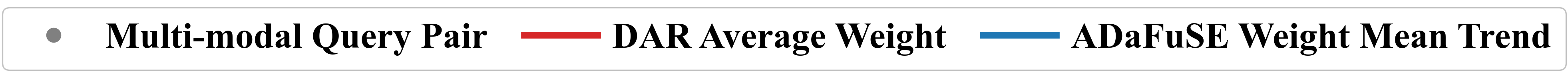}
    \end{subfigure}
    % \vspace{-1mm} 
    \caption{Modality-weighting comparison between DAR and ADaFuSE on the DA-VisDial dataset~\cite{zhang2026eliminating}.}
    \label{fig:weight}
    \vspace{-4mm}
\end{figure}

\vspace{-0.5em}
\subsection{Reduced Performance Degradation}
\label{analysis-adr}
\looseness -1 Following our analysis of the performance degradation induced by DAR's static fusion, we use the two metrics defined in Section~\ref{sec:limits} to further investigate the ability of ADaFuSE to prevent performance degradation caused by noisy generated images.
Returning to Figure~\ref{fig:adr_metric}, in sharp contrast to DAR's performance, ADaFuSE (blue solid line) consistently maintains a lower degradation rate, which further reduces as the dialogue context becomes larger. Moreover, we can observe from Figure~\ref{fig:adr_metric} (blue dashed line) that ADaFuSE markedly reduces the negative impact of noisy images, with the average rank drop for instances degraded by including the image averaging around 20 ranks, down from around 7,500 for DAR. This confirms that ADaFuSE effectively identifies and suppresses the conflicting noise introduced by diffusion models, ensuring robust retrieval performance even when diffusion-generated images are unreliable.

\vspace{-0.5em}
\subsection{Why Adaptive Gating Helps?}
\looseness -1 To illustrate the impact of the proposed adaptive gating mechanism, we visualize the assigned image weight ($1 - \lambda_{n,i}$) as defined in Eq~\ref{lambda} against the cosine similarity between the textual query and corresponding generated image in Figure~\ref{fig:weight}. In the plot, gray points represent 100k individual text-image pairs sampled from the DA-VisDial dataset~\cite{zhang2026eliminating}, the blue line depicts the fitted regression trend for ADaFuSE and the red line represents the average weight across ten rounds of the DAR. We observe that as the semantic alignment (cosine similarity) between the text and generated image feature increases, ADaFuSE adaptively amplifies the contribution of the generated image. We also see that on average, ADaFuSE is more conservative in its \ignore{(initial)} use of the generated image evidence--keeping the gating weight of the generated image between 15-25\%, in contrast to 45\% for DAR. Although note that aspects of the generated image can then be emphasized in ADaFuSE via the subsequent semantic-aware mixture-of-experts component.

\vspace{-0.5em}
\section{Conclusion}
\looseness -1 In this work, we propose ADaFuSE, a lightweight but robust fusion model for diffusion-augmented I-TIR, designed to mitigate the noise sensitivity of static additive fusion. By combining an adaptive gating mechanism with a semantic-aware mixture-of-experts component, ADaFuSE dynamically controls how much emphasis to place on the diffusion-generated images. This reduces the impact of generated noise, while preserving fine-grained semantic cues during query embedding. Extensive experiments across four benchmarks demonstrate that ADaFuSE achieves state-of-the-art performance. Moreover, our analysis highlights how ADaFuSE improves robustness across queries, as it minimizes performance degradation even when dealing with noisy generative feedback, offering a promising path toward more reliable interactive retrieval systems.

% \clearpage

\printbibliography

\end{document}